\documentclass{article}

\usepackage{amsmath}
\usepackage{amssymb}
\usepackage{cancel}
\usepackage{graphicx}
\usepackage{fullpage}
\usepackage{authblk}
\usepackage{cite}
\usepackage[dvipsnames]{xcolor}

 \newcommand{\add}[1]{#1}
 \newcommand{\remove}[1]{}
\usepackage{hyperref}

\title{Unfrustrated Self-Morphing of Bulk Liquid Crystal Elastomers}
\author[1,2]{Shachaf Rotem}
\author[1]{Hillel Aharoni \thanks{hillel.aharoni@weizmann.ac.il}$^{,}$}
\date{}
\affil[1]{Department of Complex Systems, Weizmann Institute of Science, Rehovot 7610001, Israel}
\affil[2]{School of Physics and Astronomy, Tel Aviv University, Tel Aviv 6997801, Israel}

\begin{document}
\maketitle

\begin{abstract}
Precise manipulation of shape-morphing responsive materials is crucial for applications in soft robotics and adaptive structures. While notable precision has been achieved in thin two-dimensional sheets, an accurate volumetric shape-morphing remains a major challenge due to geometric frustration, which inevitably generates complex, residual elastic stresses. In this work, we extend the geometric approach used for thin sheets to bulk Liquid Crystal Elastomers (LCEs). By examining their reference Ricci curvature, we formulate the minimal set of conditions required for a three-dimensional nematic director field to undergo stress-free, frustration-free deformations upon actuation. Through this mathematical framework, we \remove{identify two distinct classes of geometrically compatible bulk systems}\add{prove that such a director field in any open three-dimensional region is strictly required to be twistless ($T \equiv 0$)}. The \remove{first class comprises}\add{resulting family of} twistless director fields \remove{that}\add{remains} frustration-free across all temperatures, leading to holographic design principles demonstrated through ``Planar'' and ``Smectic'' LCE subfamilies. \remove{The second class features twisted}\add{Conversely, twisted} configurations \remove{that}\add{cannot achieve volumetric compatibility, but} exhibit unique, temperature-selective \add{pointwise} compatibility, leading to non-monotonic accumulation of internal elastic stresses that \remove{relax completely}\add{suppress leading-order elastic frustration in sufficiently small bodies} at a predefined target temperature. Our framework establishes a firm mathematical foundation for robust forward and inverse design protocols in bulk LCEs.
\end{abstract}

\section*{Introduction}

In recent years, there has been a significant surge of scientific and technological interest in soft robotics, adaptive structures, and functional materials \cite{yasa2023overview, walther2020responsive, xia2022responsive}, which hold immense potential for applications spanning biomedical devices, aerospace engineering, and smart architecture \cite{cianchetti2018biomedical, zheng2025engineering, zhang2023progress, kallayil2025adaptive}. These rapidly growing fields inherently require the precise design and manipulation of shape-morphing responsive materials to achieve accurate, reliable, and robust mechanical deformations \cite{zheng2025engineering, gao2022three}. Many previous works have focused on the self-morphing of thin surfaces and two-dimensional sheets, achieving remarkably high-precision capabilities in both theoretical design and physical manufacturing \cite{klein2007shaping, jeon2017shape, ware2015voxelated}. However, thin surfaces can exert only limited mechanical work and are therefore insufficient for a wide variety of practical purposes that require true bulk deformation. Recently, major technological advances in additive manufacturing and 4D printing have addressed many of the historical challenges in the accurate manufacturing of bulk-responsive materials \cite{chen2023recent, ding20254d, zhou2024additive, yarali20244d}. While these techniques enable the creation of volumetric structures with precise local responses, they introduce a profound challenge to the design problem of bulk systems. In the absence of a spare spatial dimension to buckle into, a generic uncorrelated local response will inevitably incur geometric frustration \cite{grason2016perspective, meiri2021cumulative}. Consequently, the system will develop residual elastic stresses whose effects are highly complex and difficult to control or manipulate.

Liquid Crystal Elastomers (LCEs) are a prominent class of responsive soft materials that seamlessly integrate an elastic polymer network with anisotropic liquid crystal mesogens \cite{warner2007liquid}. LCEs are widely celebrated for exhibiting a robust local mechanical response, fast actuation capabilities, and high work density when exposed to a variety of external stimuli, such as heat, light, and electric fields \cite{ohm2010liquid, terentjev2025liquid}. Previous works extensively explored the shape-morphing of thin LCE sheets\add{ with pre-programmed director fields}, often utilizing a rigorous geometric approach to map planar director fields to the resulting 3D topographies \cite{modes2011gaussian, aharoni2014geometry, ware2015voxelated, aharoni2018universal, griniasty2019curved}. As for bulk \remove{systems,}\add{LCEs, the interplay between nematic alignment and finite elasticity has long been explored in the foundational literature on soft and semi-soft elasticity \cite{DeSimone2000material, DeSimone2002macroscopic, Conti2002soft, Conti2002semi, Biggins2009supersoft, Biggins2012elasticity, Lee2023universal}. Although these works treat mechanical loading where the director is free to reorient, whereas in programmable shape morphing the reference director field is locked into the network during synthesis, the two problems coincide at the level of smooth zero-energy states: any smooth, zero-energy director configuration of an ideal soft LCE corresponds exactly to a compatible, stress-free actuated configuration of a pre-programmed locked-director LCE, and vice versa. However, in the soft LCE setting, compatibility is typically achieved through quasiconvex relaxation by forming rapidly fluctuating, non-smooth laminated microstructures (stripe domains).} Recent advances in the additive manufacturing of LCEs have successfully enabled the fabrication of bulk LCE materials with full 3D spatial control over the nematic director alignment \cite{chen2023recent, guo2021shape, herman2024digital, telles2025spatially, gulati2026holographic}. \add{For programmed shape morphing using such systems, translating non-smooth microstructures obtained from quasiconvex relaxation into a manufacturing blueprint would require infinite spatial resolution, which is bounded practically by printing voxel size and fundamentally by the nematic correlation length. At any finite manufacturing resolution, truncating director oscillations inevitably generates strong residual stresses at the voxel scale, whose macroscopic effects depend sensitively on higher-order gradients and on the specific material's elastic energy functional. Smooth inhomogeneous solutions have also been examined via Ericksen's approach to universal deformations \cite{Lee2023universal}, but they focus on specific symmetric deformation families rather than the general geometric criteria for compatibility. Thus,} the full potential of \remove{these}\add{bulk LCE manufacturing} capabilities remains largely unknown due to a lack of a theory to translate them into useful design principles.

\begin{figure}[ht]
\centering
    \centering
	\includegraphics[width=.9\linewidth]{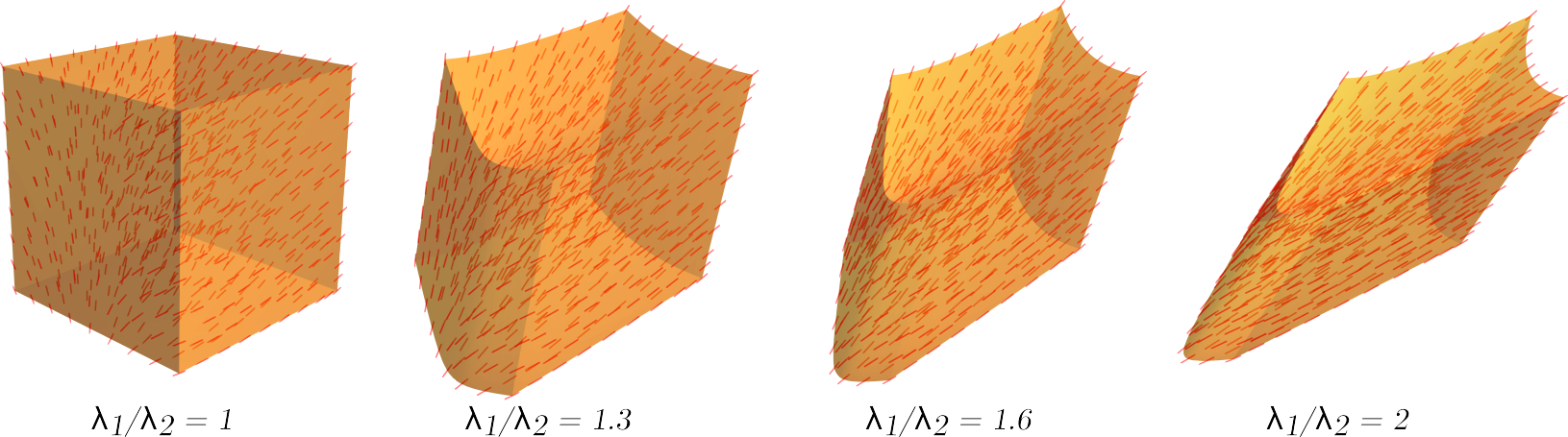}
    \caption{
    Deformation of a compatible liquid crystal elastomer (LCE) bulk. As each small element elongates by a factor $\lambda_1$ along the local director (red segments) and by a factor $\lambda_2$ in the perpendicular local directions, the LCE changes its global shape. Such an object will be elastically frustrated and develop residual stresses, unless it is endowed with a director field which induces a Ricci-flat reference metric, in which case its deformation can be fully resolved from geometric principles. \add{The depicted example belongs to the twist-free all-temperature compatible class (specifically, to the Smectic LCE subfamily), and all four configurations shown ($\lambda_1/\lambda_2 = 1, 1.3, 1.6, 2$, as well as all intermediate states) are completely stress-free Euclidean embeddings.} There is a rich family of such fields, allowing a wide variety of nontrivial shape deformations.}
    \label{fig: lambda illustration}
\end{figure}

In this work, we expand the geometric approach of thin LCE sheets to bulk LCEs. We study the family of 3D director fields for a bulk LCE that, upon actuation by a control parameter (e.g., temperature), will undergo a deformation free of geometric frustration \add{(Fig.~\ref{fig: lambda illustration})}\remove{fig. (\ref{fig: lambda illustration})}. Namely, we study LCEs whose actuated intrinsic geometry admits a stress-free configuration in ambient Euclidean space. Unlike frustrated configurations, the shape of an unfrustrated state can be determined exactly from local responses, without the need to optimize a complex (possibly anisotropic) elastic energy functional. This poses a firm mathematical basis for robust design and inverse-design protocols. We formulate the minimal set of equations required for a bulk LCE director field to remain frustration-free upon actuation\remove{, and through this framework, we identify two broad families of such director fields}\add{. Notably, we prove a general rigidity theorem showing that volumetric compatibility throughout any open 3D region strictly requires the director to be twist-free ($T \equiv 0$). For this unique class of volumetrically unfrustrated bulk LCEs, compatibility is guaranteed at all temperatures. We construct two broad holographic subfamilies of such fields}. We further utilize our approach to generate \add{twisted} director fields that exhibit a unique non-monotonic geometric frustration as a function of the deformation parameter\add{, suppressing leading-order frustration at a target temperature in small bodies}. We discuss the physical outcomes and promising technological applications of our findings.

\section*{Theoretical background}

Following numerous previous works, we model an LCE as an elastic solid with a varying director field, ${\bf n}(x,y,z)$, embedded within it via some manufacturing process. We neglect variations in the magnitude of the nematic ordering. We represent ${\bf n}$ as a vector with components $n_{i}$, where $i=1,2,3$, but recall that it represents a head-tail symmetric order of constant magnitude so that ${\bf n}\sim-{\bf n}$ and ${\bf n}\cdot{\bf n}=1$. At room temperature, the solid resides stress-free in Euclidean 3-dimensional space, and we use its $(x,y,z)$ Cartesian coordinates in this state as material (Lagrangian) coordinates. When actuated, material elements tend to elongate by a factor of $\lambda_1$ along the local director, and by a factor of $\lambda_2$ perpendicular to the director. This endows the solid with a \emph{reference metric} tensor
\begin{equation}\label{eqn: metric tensor}
    g_{ij}=\left(\lambda_{1}^{2}-\lambda_{2}^{2}\right)n_{i}n_{j}+\lambda_{2}^{2}\delta_{ij}.
\end{equation}
Any $\lambda_1 \ne \lambda_2$ reflects a local anisotropic deformation, and $\sqrt{\det g}=\lambda_1\lambda_2^2$ is the (homogeneous) volume expansion factor. The reference metric \eqref{eqn: metric tensor} can be used to calculate the \emph{reference Ricci curvature} tensor
\begin{equation} \label{eqn: definition ricci tensor}
    R_{ij}=\partial_{k}\Gamma_{ij}^{k}-\partial_{j}\Gamma_{ik}^{k}+\Gamma_{ij}^{r}\Gamma_{rk}^{k}-\Gamma_{ik}^{r}\Gamma_{rj}^{k},
\end{equation}
where $\Gamma^i_{jk}$ are the Christoffel symbols \cite{DoCarmo}
\begin{equation} \label{eqn: definition christoffel symbol}
    \Gamma_{kl}^{i}=\frac{1}{2}g^{im}\left(\partial_{l}g_{mk}+\partial_{k}g_{ml}-\partial_{m}g_{kl}\right).
\end{equation}

A deformed simply-connected elastic body may be smoothly immersed in Euclidean space strain-free, and therefore stress-free, if and only if its reference Ricci curvature ${\bf R}$ vanishes at every point \cite{willmore1956compact, efrati2013metric}. \add{More generally, one must distinguish between: (i) pointwise compatibility ($R_{ij}(p) = 0$), which suppresses leading-order elastic frustration in small neighborhoods of $p$; (ii) local volumetric compatibility ($R_{ij} \equiv 0$ in an open 3D region), which by the fundamental theorem of Riemannian geometry guarantees a local isometric immersion into $\mathbb{R}^3$; and (iii) global immersion throughout a simply connected body which requires (ii) at every point. }\remove{This observation}\add{The crucial role of ${\bf R}$ in these problems} is in the essence of the analysis we present in this paper. We look for LCEs that may be fully relieved of elastic stresses only by examining their reference Ricci curvatures.
To facilitate analyzing the reference Ricci flatness of the deformed LCE, we use the irreducible representation of the director's gradient given in \cite{selinger2022director}:
\begin{equation} \label{eqn: definition gradient n}
    \partial_{i}n_{j}=-n_{i}B_{j}+\frac{1}{2}S\left(\delta_{ij}-n_{i}n_{j}\right)+\frac{1}{2}T\epsilon_{ijk}n_{k}+\Delta_{ij}
\end{equation}
Here, $S=\nabla \cdot {\bf n}$ is the splay scalar, $\boldsymbol{B}={\bf n} \times \left(\nabla \times {\bf n}\right)$ is the bend vector (which is perpendicular to the director), $T={\bf n} \cdot \left(\nabla \times {\bf n}\right)$ is the twist scalar, and $\boldsymbol{\Delta}$ is the biaxial splay tensor, which is symmetric, traceless and perpendicular to the director.

\section*{Ricci tensor of a bulk LCE}

A direct calculation of (\ref{eqn: definition ricci tensor}), the Ricci curvature tensor of a bulk LCE with a prescribed director field ${\bf n}(x,y,z)$, results in:
\begin{equation} \label{result: ricci tensor}
	\begin{split}
		R_{ij}=(\Lambda^2-1)\biggl[&\left(-\partial_{k}B_k+\frac{\Lambda^2+1}{2}\,T^{2}\right)n_{i}n_{j}\\
		+&\frac{1}{2\Lambda^2}\left(\partial_{k}(n_kS) - \Lambda^2 T^2\right)\left(\delta_{ij}-n_{i}n_{j}\right)\\
		+& \frac{1}{\Lambda^{2}}\left(\partial_{n}\Delta_{sr}+S\Delta_{sr}-T\epsilon_{rkl}n_{k}\Delta_{sl}\right)\left(\delta_{jr}-n_{j}n_{r}\right)\left(\delta_{is}-n_{i}n_{s}\right)\\
		+& \frac{1}{2}\left(\partial_{s}T+2TB_{s}\right)\left(n_{i}n_{l}\epsilon_{jsl}+n_{j}n_{l}\epsilon_{isl}\right)\biggr],
	\end{split}
\end{equation}
where we denote $\Lambda=\frac{\lambda_1}{\lambda_2}$ and $\partial_n \equiv n_i \partial_i$ is a directional derivative along the director. A detailed derivation is presented in Appendix~\ref{sec: calculating ricci tensor}.
The four rows in eq.~\eqref{result: ricci tensor} correspond to four independent components of the Ricci tensor. Respectively,
\begin{equation}\label{result: schematic ricci tensor}
    {\bf R}=(\Lambda^2-1)\left[{\bf R}_{nn}+{\bf R}_{\perp\perp}^\textrm{trace}+{\bf R}_{\perp\perp}^\textrm{traceless}+{\bf R}_{n\perp}\right].
\end{equation}
Therefore, the Ricci curvature vanishes if and only if either $\Lambda=1$ or all four summands in eq.~\eqref{result: schematic ricci tensor} vanish simultaneously, thus:
\begin{subequations}\label{eqn: Ricci Components}
    \begin{align}
        \label{eqn: Ricci Components 1}
        -\nabla\cdot\boldsymbol{B}+\frac{\Lambda^2+1}{2}\,T^{2}&=0,\\
        \label{eqn: Ricci Components 2}
        \nabla\cdot\left({\bf n}S\right)-\Lambda^2\,T^{2}&=0,\\
        \label{eqn: Ricci Components 3}
        \mathcal{P}_\perp\left(\partial_{n}\boldsymbol{\Delta}+S\boldsymbol{\Delta}-T{\bf n}\times\boldsymbol{\Delta}\right)&=0\\
        \label{eqn: Ricci Components 4}
        \nabla_{\perp}T+2T\boldsymbol{B}&=0,
    \end{align}
\end{subequations}
where $\mathcal{P_\perp}$ is a projection operator onto the subspace perpendicular to ${\bf n}$ and $\nabla_{\perp}[\cdot]=\mathcal{P}_{\perp} \nabla[\cdot]$ is the projection of the gradient onto that subspace.
Equations~\eqref{eqn: Ricci Components 1} and \eqref{eqn: Ricci Components 2} are scalar, while eq.~\eqref{eqn: Ricci Components 3} and \eqref{eqn: Ricci Components 4} have 2 scalar degrees of freedom each (representing a 2D-symmetric-traceless tensor and a 2D vector perpendicular to ${\bf n}$, respectively). 
Together, they account for all 6 degrees of freedom in the Ricci curvature tensor. Equations~\eqref{eqn: Ricci Components} can be further simplified (Appendix~\ref{sec: refining compatibility conditions}) to read
\begin{subequations}\label{eqn: Simplified Ricci} 
    \begin{gather}
		\label{eqn: Simplified Ricci 1}
		\Lambda^2\,T^2~=~\frac{2}{3+\Lambda^{-2}}\,\sigma~=~\nabla\cdot({\bf n} S),\\ 
		\label{eqn: Simplified Ricci 2}          
        T=T_{\Delta},\\
        \label{eqn: Simplified Ricci 3}
        S=-\partial_{n}(\log\Delta),\\
        \label{eqn: Simplified Ricci 4}
        \nabla_{\perp}T=-2\,T\boldsymbol{B}.  
    \end{gather}
\end{subequations}
Here we introduced the well-known saddle-splay, $\sigma$, defined \cite{selinger2019interpretation} as
\begin{equation} \label{def: saddle-splay}
    \sigma\equiv \nabla \cdot ({\bf n} S+\boldsymbol{B})=\frac{1}{2} S^2+\frac{1}{2}T^2-2\Delta^2
\end{equation}
and a new quantity we call the \emph{$\Delta$-twist}, $T_{\Delta}$, defined \add{for $\Delta \neq 0$} as
\begin{equation} \label{def: delta-twist}
    T_{\Delta}=\tfrac{1}{2\Delta^2}\mathrm{Tr}\left(({\bf n}\times\boldsymbol{\Delta})\cdot\partial_n\boldsymbol{\Delta}\right).
\end{equation}
In the above notation, we introduced $\Delta\ge0$ the scalar magnitude of $\boldsymbol{\Delta}$, namely $\boldsymbol{\Delta}$ assumes eigenvalues $(0,\Delta,-\Delta)$ matching an orthonormal eigenvector triad $({\bf n},{\bf e},{\bf n}\times{\bf e})$, respectively (equivalently, define $\Delta=\sqrt{\tfrac{1}{2}\mathrm{Tr}(\boldsymbol{\Delta^2})}$). The physical essence of $T_{\Delta}$ is made clear by rewriting \eqref{def: delta-twist} in the form $T_{\Delta}=2({\bf n}\times{\bf e})\cdot\partial_n{\bf e}$, thus, the $\Delta$-twist $T_\Delta$ is the rate at which the $\boldsymbol{\Delta}$ rotates within the plane perpendicular to the director, when moving along the director. \add{Note that Eqs.~\eqref{eqn: Simplified Ricci 2} and \eqref{eqn: Simplified Ricci 3} are obtained from Eq.~\eqref{eqn: Ricci Components 3} by projection under the assumption $\Delta \neq 0$; at points where $\Delta = 0$, Eq.~\eqref{eqn: Ricci Components 3} is governed directly by $\mathcal{P}_\perp(\partial_n \boldsymbol{\Delta}) = 0$ without division.}

Notably, only equation~\eqref{eqn: Simplified Ricci 1} depends on the temperature. This establishes a qualitative distinction between two \remove{families of geometrically compatible bulk LCEs:}\add{scenarios: (i) all-temperature volumetric compatibility, where} eq.~\eqref{eqn: Simplified Ricci 1} equals zero, making the entire equation set~\eqref{eqn: Simplified Ricci} temperature-independent; and \remove{those where}\add{(ii) temperature-selective compatibility, where} eq.~\eqref{eqn: Simplified Ricci 1} is nonzero. \add{Crucially, in Appendix~\ref{sec: appendix twist}, we prove that if the compatibility conditions \eqref{eqn: Ricci Components} hold throughout an open 3D domain for $\Lambda \neq 1$, the twist must vanish identically ($T \equiv 0$). Hence, volumetric compatibility across an extended 3D region is strictly restricted to the twistless class ($T \equiv 0$).}

The two \remove{families}\add{regimes} behave very differently. In the former \add{($T=0$)}, if the Ricci tensor vanishes at one nontrivial temperature (namely $\Lambda\neq 1$), it vanishes at all temperatures. This is similar to the established homogeneity in 2D LCEs \cite{aharoni2014geometry,mostajeran2015curvature}. Thus, when a bulk LCE in this family is heated or cooled slowly, it will smoothly change shape while remaining stress-free throughout. In the second \remove{family}\add{case ($T \neq 0$)}, however, \add{because $\Lambda^2 = \nabla\cdot({\bf n}S)/T^2$ is uniquely determined by the reference director at any point where $T \neq 0$,} the Ricci tensor may only vanish at $\Lambda=1$ and at \remove{some other particular value of $\Lambda$}\add{at most one nontrivial target value $\Lambda_T = \sqrt{\nabla\cdot({\bf n}S)/T^2} > 0$}. A bulk LCE with twisted director cannot adopt a stress-free configuration throughout a 3D volume, and will accumulate residual stresses when heated (or cooled). However, strikingly, this stress accumulation \add{in small bodies} is non-monotonic; as one heats (or cools) to a predefined target temperature, \remove{all residual elastic stresses suddenly vanish}\add{the leading-order elastic frustration is suppressed}, leaving the LCE \remove{perfectly relaxed in a configuration different from its initial state}\add{locally unfrustrated near the origin}. We next discuss these two regimes in further depth.

\section*{All-temperature compatibility}
We start with the case where eq.~\eqref{eqn: Simplified Ricci 1} vanishes, thus
\begin{equation} \label{eqn: necessary condition zero}
	T=T_{\Delta}=\sigma=\nabla\cdot({\bf n}\,S)=0.
\end{equation}

Eqs.~\eqref{eqn: necessary condition zero}, should they apply in the bulk, immediately imply eq.~\eqref{eqn: Simplified Ricci 4} by the fact that $T\equiv0$ throughout the bulk. In addition, since $T=0$ and $\sigma=0$, from eq.~\eqref{def: saddle-splay} we find that $S^2=4\Delta^2$, thus $S=\pm 2\Delta$. Thus, \add{wherever $S \neq 0$ (so $\Delta = |S|/2 \neq 0$),} from eqs.~\eqref{eqn: necessary condition zero},
\begin{equation} \label{eqn: S log Delta identity}
    S+\partial_n(\log\Delta)=S+\partial_n(\log \add{|S|})=\tfrac{1}{S}(S^2+\partial_n S)=\tfrac{1}{S}(S\,\nabla\cdot{\bf n}+{\bf n}\cdot\nabla S)=\tfrac{1}{S}\nabla\cdot({\bf n}\,S)=0
\end{equation}
and therefore eq.~\eqref{eqn: Simplified Ricci 3} is also satisfied. \add{At points where $S = 0$, we have $\Delta = 0$, and Eq.~\eqref{eqn: Ricci Components 3} is satisfied identically since $\boldsymbol{\Delta} = 0$, and $\nabla\cdot({\bf n}S)=0$ implies $\partial_n S = \partial_n \Delta = 0$, so that $\partial_n \boldsymbol{\Delta} = 0$} Hence, remarkably, eqs.~\eqref{eqn: necessary condition zero} are sufficient and necessary conditions for an LCE that remains compatible at all temperatures.

\begin{figure}[ht]
    \centering
	\includegraphics[width=\linewidth]{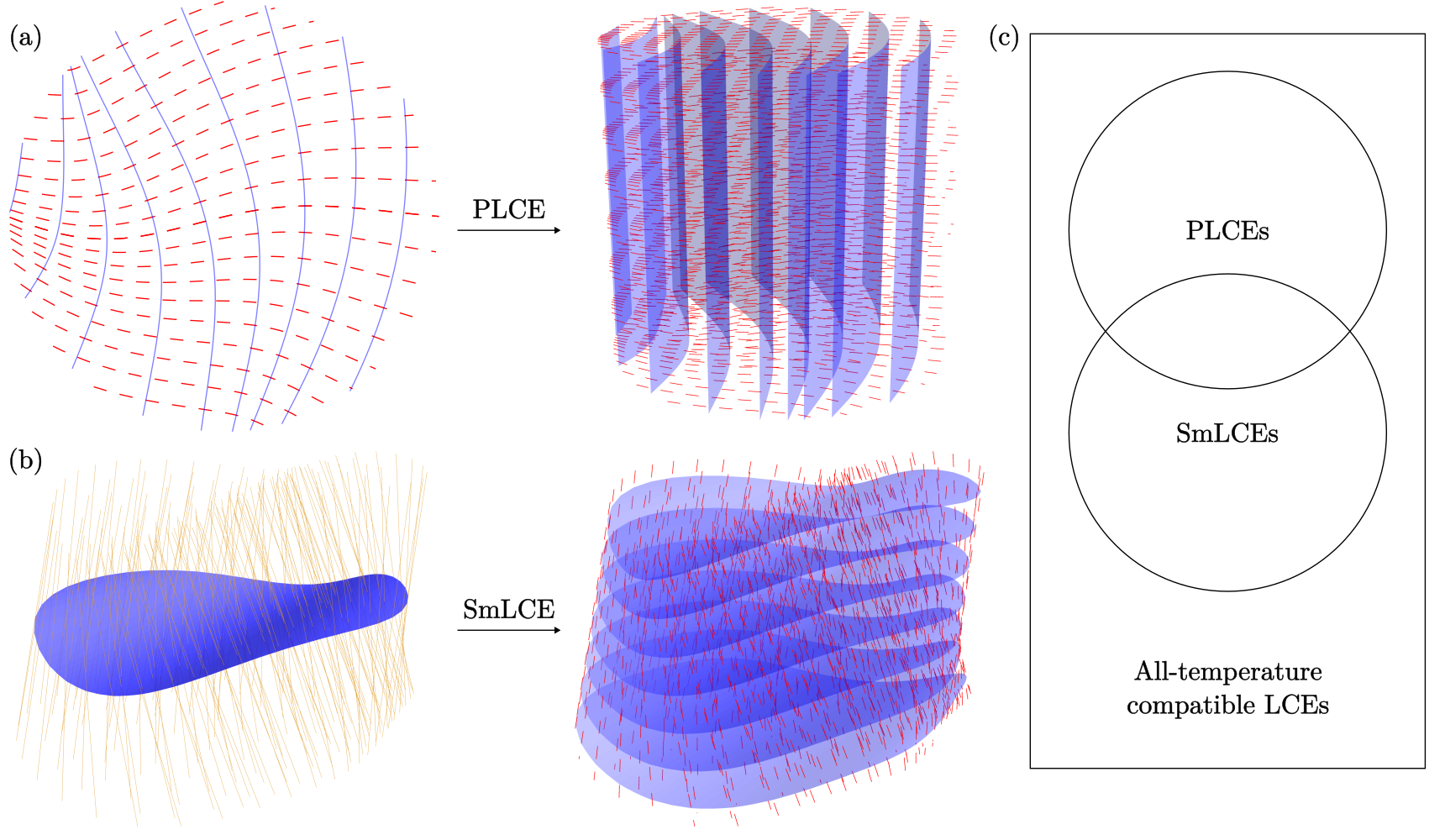}
    \caption{All-temperature compatible bulk LCEs. (a) Planar LCEs (PLCEs), introduced and fully resolved in \cite{PLCEcastro}, are a class of 2D director fields. Stacked together, they make a bulk LCE that deforms without frustration at all temperatures. The director field is shown by the red segments, and director perpendiculars by the blue surfaces. (b) A director field aligned with the normals (orange) to any developable surface (blue) yields at all temperatures a frustration-free bulk LCE with uniformly spaced layers; thus, we call it a Smectic LCE (SmLCE). (c) There are all-temperature compatible bulk LCEs that belong to neither of these two subfamilies (or to one of them, or to both).}
    \label{fig: All-temp LCEs}
\end{figure}

Director fields that satisfy eqs.~\eqref{eqn: necessary condition zero} have several interesting geometric properties: because the twist $T$ vanishes, there is a foliation of space into surfaces perpendicular to the director. The Gaussian curvature of these surfaces is \remove{the saddle-splay $\sigma$}\add{half the saddle-splay, $K = \sigma/2$}, which also vanishes, thus they are developable. The other two terms in eq.~\eqref{eqn: necessary condition zero} also have a geometric interpretation, albeit more subtle. The vanishing of $T_{\Delta}$ implies that the flat surface-direction is parallel-transported along the director across different surfaces in the foliation, and the vanishing of $\nabla\cdot({\bf n} S)$ implies that the foliation is in equilibrium of the mean curvature flow. These properties may be a signature of a more elegant structure underlying compatible director fields, but we have not found such a structure thus far. One may alternatively expect that eqs.~\eqref{eqn: necessary condition zero} are simply too restrictive, leaving only trivial director fields at their intersection. In the following subsections, we show that this is not the case by presenting two extended families of director fields that satisfy these criteria. Shown in \add{Fig.~\ref{fig: All-temp LCEs}}\remove{fig. (\ref{fig: All-temp LCEs})} are examples from both families.

\subsection*{Planar LCEs}
Planar LCEs, or PLCEs, are LCEs with 2D director fields, namely ${\bf n}(x,y,z)={\bf n}(x,y)$ and ${\bf z}\cdot{\bf n}=0$, whose Gaussian curvature vanishes at all $\Lambda$s. Like all 2D director fields they satisfy $T=0$ \cite{da2023compatible}, and it was shown in \cite{PLCEcastro} that the conditions $\nabla \cdot \boldsymbol{B} = \nabla \cdot ({\bf n} S)=0$ are necessary and sufficient to satisfy the vanishing Gaussian curvature requirement.

Furthermore, \cite{PLCEcastro} deduced that as a bulk LCE, PLCEs will remain compatible at all temperatures and, indeed, PLCEs satisfy eqs.~\eqref{eqn: necessary condition zero}: from the above, $T$, $\nabla \cdot ({\bf n} S)$, and $\sigma=\nabla \cdot ({\bf n} S + \boldsymbol{B})$ all vanish; it is easy to see that $T_\Delta$ vanishes as well by noticing that one of the nontrivial eigenvectors of $\boldsymbol{\Delta}$ is ${\bf z}$ so it is constant in space (and in particular constant along ${\bf n}$). All PLCE director fields were obtained and classified in \cite{PLCEcastro} as unique solutions to a Goursat problem in the plane. Thus, PLCEs are a rich nontrivial family of LCEs compatible at all temperatures.

\subsection*{"Smectic" LCEs}
Another family of compatible LCEs can be constructed from nematic configurations with vanishing bend and twist, namely nematic director fields admissible by a Smectic-A liquid crystal bulk \cite{degennes1993physics}\add{ (distinct from physical smectic-A elastomers where molecular layer-spacing constraints dictate soft elasticity \cite{Adams2005elasticity, Adams2007soft})}. To be exact, the conditions $\boldsymbol{B}=0$ and $T=0$ imply that ${\bf n}=\nabla \phi$ and that the scalar potential $\phi(\boldsymbol{x})$ admits equidistant level sets -- the smectic layers. Smectic director fields with $\boldsymbol{B}=0$ and $T=0$ are holographic in the sense that the entire field is determined by a single smectic layer. The field everywhere follows straight lines that intersect the initial layer perpendicularly \cite{da2023compatible}. The construction admits a finite horizon at the point where these lines intersect, namely at a distance equal to the minimal radius of curvature of the initial surface. At the horizon, the director field becomes discontinuous, at which point several of our framework's assumptions break, and a more careful treatment is required.

Smectic director fields satisfy eqs.~\eqref{eqn: necessary condition zero} if and only if we add the requirement that $\sigma=0$, namely that the Gaussian curvature of the smectic layers vanishes \add{($K = \sigma/2 = 0$)}. This condition guarantees that $\nabla\cdot({\bf n} S)=\sigma-\nabla\cdot \boldsymbol{B}=0$. In addition, it guarantees that at every point there is a flat direction of the smectic layer, which is one of the nontrivial eigenvectors of $\boldsymbol{\Delta}$. By the holographic construction, this direction is constant along ${\bf n}$, thus it implies $T_{\Delta}=0$ and concludes eqs.~\eqref{eqn: necessary condition zero} and therefore the vanishing of the entire Ricci Tensor at all temperatures.

\section*{Temperature-selective compatibility}

We now turn to discuss LCEs whose director fields satisfy eqs.~\eqref{eqn: Simplified Ricci}, however, with assigning a nonzero value to eq.~\eqref{eqn: Simplified Ricci 1}. As discussed before, the temperature dependence of eq.~\eqref{eqn: Simplified Ricci 1} implies that the Ricci tensor for those LCEs may only vanish at a specific target value of $\Lambda$, or target temperature, and not at all temperatures.

Unlike all-temperature compatible LCEs, where the vanishing of the twist greatly reduces the complexity of the Ricci tensor to eq.~\eqref{eqn: necessary condition zero}, temperature-selective compatible director fields are subject to the full, generically overdetermined, system of equations \eqref{eqn: Simplified Ricci}. \remove{While we find it unlikely that this system with $T\ne0$ could be satisfied in a nontrivial 3D volume, we have no direct proof of this conjecture.}\add{In Appendix~\ref{sec: appendix twist}, we rigorously prove that this system with $T \ne 0$ admits no smooth solution in any open 3D volume.}

It is nonetheless not hard to construct a director field that satisfies the temperature-selective compatibility conditions \emph{at a single point}. We shall demonstrate the exceptional properties of such LCE using the following concrete example:
\begin{equation}
    {\bf n}(x,y,z)=n_x\hat{x}+n_y\hat{y}+\sqrt{1-n_x^2-n_y^2}\,\hat{z},
\end{equation}
with 
\begin{equation}
\begin{split} \label{director field two temperatures}
    n_x &= \frac{-y + \sqrt{3}\Lambda_T x}{2} - \Lambda_T^2 x z,\\
	n_y &= \frac{x + \sqrt{3}\Lambda_T y}{2} - \Lambda_T^2 y z,
\end{split}
\end{equation}
where $\Lambda_T$ is some constant. \add{Here, $x, y, z$ are dimensionless Cartesian coordinates normalized by the characteristic length scale of the director variation, and $|{\bf x}| = \sqrt{x^2+y^2+z^2} \ll 1$ denotes the distance from the origin.} Then,
\begin{equation}
\begin{split}
	S &= \sqrt{3}\Lambda_T - 2\Lambda_T^2 z + O(|{\bf x}|^2),\\
	T &= 1 + O(|{\bf x}|^2),\\
    \boldsymbol{B} &= \frac{1}{4}\left( (1 + \Lambda_T^2) x + 2\sqrt{3}\Lambda_T y, (1 + \Lambda_T^2) y - 2\sqrt{3}\Lambda_T x, 0\right) + O(|{\bf x}|^2),\\
    \boldsymbol{\Delta} &= O(|{\bf x}|^2).
\end{split}
\end{equation}
Thus, from equation~\eqref{result: ricci tensor}, the Ricci tensor reads
\begin{equation} \label{Ricci tensor two temperatures}
    R_{ij}=\frac{1}{2}\left(\Lambda^2-1\right)\left(\Lambda^2-\Lambda_T^2\right)\left[n_{i}n_{j}
		-\Lambda^{-2}\left(\delta_{ij}-n_{i}n_{j}\right)\right] + O(|{\bf x}|).
\end{equation}
Thus, the entire Ricci tensor vanishes at the origin when $\Lambda=1$ (namely in the undeformed ``room temperature'' state) or when $\Lambda=\Lambda_T$, but not at any other value of $\Lambda$.

\begin{figure}[ht]
    \centering
	\includegraphics[width=.65\linewidth]{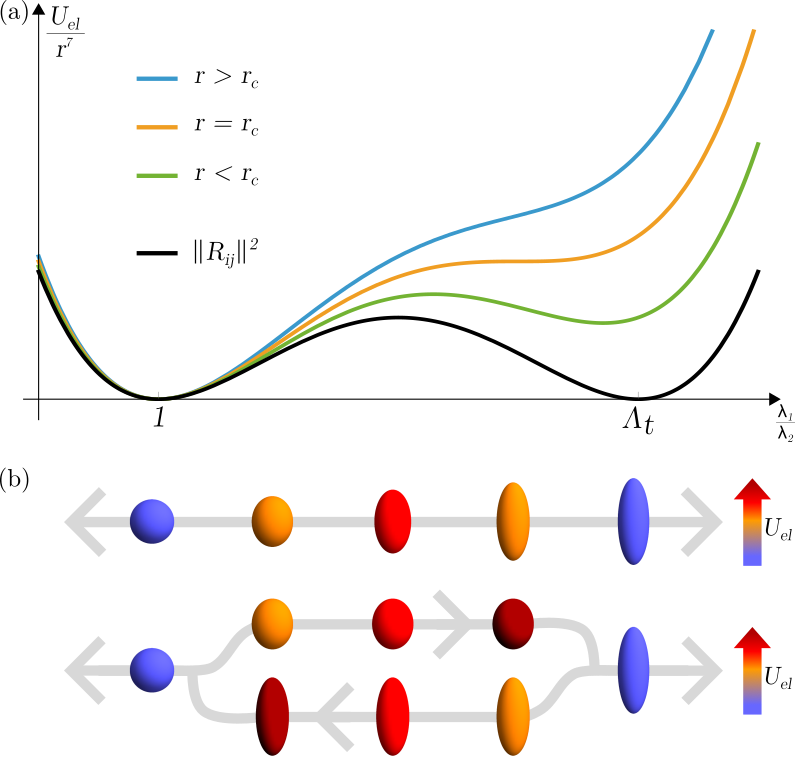}
    \caption{Temperature-selective elastic frustration.
    (a) An LCE with director field \eqref{director field two temperatures} is Ricci-flat \remove{near}\add{at} the origin only at the room- and target-temperatures, $\lambda_1/\lambda_2=1,\,\Lambda_T$, respectively\add{; the solid black curve depicts $\|R_{ij}({\bf 0})\|^2$ at the origin}. The residual stored elastic energy at equilibrium of a small LCE ball (described by eq.~\eqref{eqn: stored elastic energy lambda}) is then, for a sufficiently small ball, a non-monotonic function of the degree of local deformation. Such an object may \remove{relax}\add{suppress} its elastic stresses at room temperature and \remove{near}\add{at} the target temperature \remove{(in a different configuration)}\add{to leading order}, but not at intermediate temperatures.
    (b) A schematic description of a slow heating/cooling process. The horizontal $\lambda_1/\lambda_2$ axis is the same as (a), and the color indicates the degree of elastic frustration (residual stored energy). The process may be either reversible (top) or hysteretic (bottom), depending on details of the frustrated states and the form of the elastic moduli tensor.}
    \label{fig: ricci v.s. temperature}
\end{figure}

To demonstrate the outcomes of this result, let us consider an LCE ball of small radius $r$ with director field given by eq.~\eqref{director field two temperatures} (the origin of the coordinate system is at the center of the ball)\add{ \cite{Maghsoodi2023softening}}. The residual elastic energy of the ball at its equilibrium state is \cite{AharoniSpaceCurvature2016}
\begin{equation}\label{eqn: stored elastic energy}
    U_\textrm{el}\propto E\|{\bf R}\|^2r^7+O(r^9),
\end{equation}
where $\|R_{ij}\|^2$ is a quadratic norm of the Ricci tensor \add{at the origin}, whose exact form depends on the specific material used and may be anisotropic. We nonetheless assume that the material is characterized by a typical elastic modulus $E(\Lambda)$ (which may include a smooth non-singular dependence on $\Lambda$), and any anisotropy enters as dimensionless coefficients in $\|\cdot\|^2$. By eqs.~\eqref{Ricci tensor two temperatures} and \eqref{eqn: stored elastic energy}, we get that
\begin{equation}\label{eqn: stored elastic energy lambda}
    U_\textrm{el}\propto E(\Lambda)\left(\Lambda^2-1\right)^2\left(\Lambda^2-\Lambda_T^2\right)^2r^7+O(r^9).
\end{equation}
Thus, for $r\ll 1$ (where unity is the typical deformation length scale of the director \eqref{director field two temperatures}), the stored elastic energy assumes two local minima as a function of \remove{in }$\Lambda$ \add{(Fig.~\ref{fig: ricci v.s. temperature})}\remove{fig. (\ref{fig: ricci v.s. temperature})}. When heated from room temperature, the deformed reference metric of the LCE is non-Euclidean and is thus unable to find a stress-free configuration. However, when heated further, the even more deformed reference metric is again almost exactly Euclidean \add{near the center}. \remove{For $r\gtrsim 1$, monotonicity is restored; residual energy assumes a single minimum, since the $O(r^9)$ term and all higher-order terms in eq.~\eqref{eqn: stored elastic energy lambda} vanish at $\Lambda=1$ but not at other values. Generally, for any}\add{The $O(r^9)$ term (arising from spatial gradients $\nabla R_{ij}$ away from the origin) is strictly positive at $\Lambda=\Lambda_T\neq 1$. It lifts the second minimum to a positive residual energy $U_\textrm{el}(\Lambda_T) \propto r^9 > 0$, but for sufficiently small $r$ cannot remove it altogether. Therefore, for a small LCE ball endowed with a} director field that satisfies eqs.~\eqref{eqn: Simplified Ricci} with a nonzero value for eq.~\eqref{eqn: Simplified Ricci 1}, elastic energy at equilibrium accumulates non-monotonically with temperature. \add{Typically, unless carefully tailored otherwise, we may expect the $O(r^9)$ term to monotonically increase between the two local minima, and remove the second minimum when the radius exceeds a critical value $r_c$ (Fig.~\ref{fig: ricci v.s. temperature}). In the general case, however, confirming the disappearance of the minimum above $r_c$ requires non-asymptotic finite elasticity or numerical simulations.}

The above analysis, based solely on the reference Ricci curvature, allows estimating the elastic energy of the energy-minimizing state, but not the energetic cost of its elastic deformations at each value of $\Lambda$. The nature of the shape deformation as one slowly creeps $\Lambda$ depends on this energy landscape, with two main qualitative scenarios \add{(Fig.~\ref{fig: ricci v.s. temperature}b)}\remove{fig. (\ref{fig: ricci v.s. temperature}b)}. In one scenario, there is a single equilibrium configuration for any $\Lambda$. The sequence of equilibrium shapes as a function of $\Lambda$s is continuous and reversible. In the second scenario, intermediate values of $\Lambda$ exhibit two distinct equilibria, one near the $\Lambda=1$ equilibrium and one near the $\Lambda=\Lambda_T$ equilibrium. In this scenario, a hysteresis loop emerges upon heating and cooling the system, with a discontinuous instability-induced ``snap'' in either direction, similar to fast-actuation mechanisms used by plants or people \cite{forterre2005venus,baumgartner2020lesson}.

\section*{Discussion}

In this paper, we analyzed the Ricci tensor of a three-dimensional LCE to explore nematic director fields that undergo actuation while remaining free of mechanical frustration. Our findings suggest that nematic twist plays a core role in the behavior of these objects by separating the two families of geometrically compatible LCEs. The first family consists of twistless director fields that satisfy eqs.~\eqref{eqn: necessary condition zero}. Their compatibility does not depend on the expansion coefficients, and they exhibit smooth, parameter-controlled shape deformation completely free of elastic stresses. The second family consists of twisted nematic directors. Their compatibility conditions are satisfied only at specific values of the expansion coefficients, and \remove{(by conjecture) only on a lower-dimensional submanifold and not in a finite volume}\add{as we proved in Appendix~\ref{sec: appendix twist}, strictly on lower-dimensional submanifolds (such as a point) and never throughout an open 3D volume}. They can be used to make LCEs whose elastic frustration is minimal at two distinct values of the control parameter, with either a reversible or a hysteretic transition between them. In this work, we focused on the fundamental geometrical properties of compatible LCEs and highlighted general design principles that emerge from them. We hope that our work is found useful in designing specific experimental and technological implementations.

\add{Our framework also sheds light on the role of topological defects and singularities of the metric tensor. In 2D thin LCE sheets, isolated disclinations concentrate singular Gaussian curvature at their cores while maintaining flatness elsewhere, generating conical and anticonical buckled morphologies \cite{modes2011gaussian}. In 3D bulk LCEs, compatible twist-free configurations generically encounter focal singularities (caustics), as seen in the finite horizon of SmLCEs where normal rays intersect. More generally, bulk LCEs programmed with line or point defects can maintain $R_{ij} = 0$ throughout the bulk away from the defect cores. If these singular cores are excised (producing hollow or topologically non-trivial actuators), non-trivial holonomy allows large, global stress-free morphing without requiring local compatibility at the singularity.}

As widely argued in the literature \cite{aharoni2018universal,jiao2023mechanical,deng2022inverse}, a major goal in the study of self-morphing systems is to obtain inverse design capabilities, namely the ability to design and manufacture a system that exhibits a desired deformation. Full inverse design of deformations in 3D is not within our current reach; however, the simple geometric relations \eqref{eqn: necessary condition zero} that govern all-temperature compatible LCEs are a significant step toward this goal. In the two fully resolved subfamilies of solutions that we have identified, PLCEs and SmLCEs, we find an initial value problem that translates to a holographic design principle; controlling the deformation along a 2D developable surface uniquely defines a bulk LCE around the control surface (the design principles for PLCEs are outlined in \cite{PLCEcastro}). Indeed, for many applications, the concern is only the exact deformation at the boundaries (through which an actuator or a soft robot would interact with its surroundings), while the consistent bulk deformation provides shape robustness and mechanical strength.

The two all-temperature-compatible families of LCEs we discussed here are also useful from a manufacturing perspective, as they exhibit properties that relax the generic requirement for full voxel-by-voxel control of the LCE's 3D director. For PLCEs, the director is planar, and control over the tilt angle is not required. Furthermore, since all parallel planes are identical, plane-by-plane methods could be considered. Not surprisingly, SmLCEs could be easily realized using smectic-A liquid crystals. Put in a cell with a developable boundary of a desirable shape, the medium will arrange itself such that its nematic director is within our SmLCE family, at which point the medium can be solidified (polymerized and crosslinked).

Lastly, a more careful treatment is needed to characterize the dynamic behavior of LCEs from the selective-temperature-compatibility (twisted) family. The possibility of a hysteretic snapping mechanism combined with an accurate shape control at the target temperatures is unique and opens the door to many possibilities. Furthermore, an additional control parameter (e.g., a hard boundary) can induce a bifurcation between the reversible and hysteretic regimes. These features enable the design of complex, accurately controlled snap mechanisms. 

\section*{Acknowledgments}
This research was supported by the Israel Science Foundation (Grant No. 2677/20).

%\bibliographystyle{unsrt}
%\bibliography{bibliography.bib}

\appendix
\numberwithin{equation}{section}

\newpage
\section{Calculating Ricci tensor}\label{sec: calculating ricci tensor}

\subsection{Calculating the Christoffel symbol} \label{sec: calculating christoffel symbol}
To calculate $\Gamma_{kl}^{i}$, we note that:
\begin{equation*}
	\partial_{i}g_{jk}=\left(\lambda_{1}^{2}-\lambda_{2}^{2}\right)\partial_{i}\left(n_{j}n_{k}\right)=\left(\lambda_{1}^{2}-\lambda_{2}^{2}\right)\left(n_{j}\partial_{i}n_{k}+n_{k}\partial_{i}n_{j}\right)
\end{equation*}

\begin{eqnarray*}
	\frac{\partial_{l}g_{mk}+\partial_{k}g_{ml}-\partial_{m}g_{kl}}{\lambda_{1}^{2}-\lambda_{2}^{2}}&=&n_{m}\partial_{l}n_{k}+n_{k}\partial_{l}n_{m}+n_{m}\partial_{k}n_{l}+n_{l}\partial_{k}n_{m}-\left(n_{k}\partial_{m}n_{l}+n_{l}\partial_{m}n_{k}\right)\\&=&n_{k}\left(\partial_{l}n_{m}-\partial_{m}n_{l}\right)+n_{m}\left(\partial_{l}n_{k}+\partial_{k}n_{l}\right)+n_{l}\left(\partial_{k}n_{m}-\partial_{m}n_{k}\right)
\end{eqnarray*}
We use \eqref{eqn: definition gradient n}, and plug this into the calculation of $\Gamma^i_{kl}$:
\begin{eqnarray*}
	\Gamma_{kl}^{i}&=&\frac{1}{2}\left(\lambda_{1}^{2}-\lambda_{2}^{2}\right)g^{im}\left[n_{k}\left(\partial_{l}n_{m}-\partial_{m}n_{l}\right)+n_{m}\left(\partial_{l}n_{k}+\partial_{k}n_{l}\right)+n_{l}\left(\partial_{k}n_{m}-\partial_{m}n_{k}\right)\right]\\&=&\frac{1}{2}\left(\lambda_{1}^{2}-\lambda_{2}^{2}\right)\left[\left(\frac{1}{\lambda_{1}^{2}}-\frac{1}{\lambda_{2}^{2}}\right)n_{i}n_{m}+\frac{1}{\lambda_{2}^{2}}\delta_{im}\right]\left[n_{k}\left(\partial_{l}n_{m}-\partial_{m}n_{l}\right)+n_{m}\left(\partial_{l}n_{k}+\partial_{k}n_{l}\right)+n_{l}\left(\partial_{k}n_{m}-\partial_{m}n_{k}\right)\right]\\&&\\&=&\frac{1}{2}\left(1-\frac{\lambda_{2}^{2}}{\lambda_{1}^{2}}\right)n^{i}\left[S\left(\delta_{lk}-n_{l}n_{k}\right)+2\Delta_{lk}-\left(n_{l}B_{k}+n_{k}B_{l}\right)\right]\\&+&\frac{1}{2}\left(\lambda_{1}^{2}-\lambda_{2}^{2}\right)n_{k}\left[\frac{1}{\lambda_{1}^{2}}n_{i}B_{l}+\frac{1}{\lambda_{2}^{2}}\left(-n_{l}B_{i}+T\epsilon_{lir}n_{r}\right)\right]\\&+&\frac{1}{2}\left(\lambda_{1}^{2}-\lambda_{2}^{2}\right)n_{l}\left[\frac{1}{\lambda_{1}^{2}}n_{i}B_{k}+\frac{1}{\lambda_{2}^{2}}\left(-n_{k}B_{i}+T\epsilon_{kir}n_{r}\right)\right]\\&&\\&=&\frac{1}{2}\left(1-\frac{\lambda_{2}^{2}}{\lambda_{1}^{2}}\right)n^{i}\left[S\left(\delta_{lk}-n_{l}n_{k}\right)+2\Delta_{lk}-\cancel{\left(n_{l}B_{k}+n_{k}B_{l}\right)}\right]\\&+&\frac{1}{2}\left(\lambda_{1}^{2}-\lambda_{2}^{2}\right)\left(-\frac{2}{\lambda_{2}^{2}}n_{k}n_{l}B_{i}+\frac{T}{\lambda_{2}^{2}}n_{r}\left(n_{k}\epsilon_{lir}+n_{l}\epsilon_{kir}\right)\right)+\cancel{\frac{1}{2}\left(1-\frac{\lambda_{2}^{2}}{\lambda_{1}^{2}}\right)n_{i}\left(n_{l}B_{k}+n_{k}B_{l}\right)}\\&=&\frac{1}{2}\left(1-\frac{\lambda_{2}^{2}}{\lambda_{1}^{2}}\right)n^{i}\left[S\left(\delta_{lk}-n_{l}n_{k}\right)+2\Delta_{lk}\right]+\frac{1}{2}\left(\frac{\lambda_{1}^{2}}{\lambda_{2}^{2}}-1\right)\left(-2n_{k}n_{l}B_{i}+Tn_{r}\left(n_{k}\epsilon_{lir}+n_{l}\epsilon_{kir}\right)\right)
\end{eqnarray*}
Where we used the identities
\begin{equation*}
	\begin{split}
		\partial_{i}n_{j}-\partial_{j}n_{i}&=-n_{i}B_{j}+\frac{1}{2}S\left(\delta_{ij}-n_{i}n_{j}\right)+\frac{1}{2}T\epsilon_{ijk}n_{k}+\Delta_{ij}-\left(-n_{j}B_{i}+\frac{1}{2}S\left(\delta_{ij}-n_{i}n_{j}\right)+\frac{1}{2}T\epsilon_{jik}n_{k}+\Delta_{ij}\right)\\
		&=-\left(n_{i}B_{j}-n_{j}B_{i}\right)+T\epsilon_{ijk}n_{k}
	\end{split}
\end{equation*}
\begin{equation*}
	\begin{split}
		\partial_{i}n_{j}+\partial_{j}n_{i}&=-n_{i}B_{j}+\frac{1}{2}S\left(\delta_{ij}-n_{i}n_{j}\right)+\frac{1}{2}T\epsilon_{ijk}n_{k}+\Delta_{ij}-n_{j}B_{i}+\frac{1}{2}S\left(\delta_{ij}-n_{i}n_{j}\right)+\frac{1}{2}T\epsilon_{jik}n_{k}+\Delta_{ij}\\
		&=-\left(n_{i}B_{j}+n_{j}B_{i}\right)+S\left(\delta_{ij}-n_{i}n_{j}\right)+2\Delta_{ij}
	\end{split}
\end{equation*}
So our result for Christoffel symbol is
\begin{equation*} \label{eqn: Christoffel symbol result}
	\Gamma_{kl}^{i}=\left(1-\frac{\lambda_{2}^{2}}{\lambda_{1}^{2}}\right)n_{i}\left[\frac{1}{2}S\left(\delta_{lk}-n_{l}n_{k}\right)+\Delta_{lk}\right]+\left(\frac{\lambda_{1}^{2}}{\lambda_{2}^{2}}-1\right)\left(-n_{k}n_{l}B_{i}+\frac{1}{2}Tn_{r}\left(n_{k}\epsilon_{lir}+n_{l}\epsilon_{kir}\right)\right)
\end{equation*}

\subsection{Calculating the Ricci tensor}

We will calculate each of the terms of eq. \eqref{eqn: definition ricci tensor} separately. For brevity we'll denote $\Lambda_1 =\frac{\lambda_1^2-\lambda_2^2}{\lambda_1^2}$ and $\Lambda_2 =\frac{\lambda_1^2-\lambda_2^2}{\lambda_2^2}$.  
\begin{equation*}
	\begin{split}
		\partial_{k}\Gamma_{ij}^{k}
		=&\partial_{s}\left[\Lambda_{1}n_{s}\partial_{i}n_{j}+\Lambda_{1}n_{s}n_{i}B_{j}-\Lambda_{2}n_{i}n_{j}B_{s}+\frac{1}{2}Tn_{r}\left(\Lambda_{2}n_{i}\epsilon_{jsr}+\Lambda_{2}n_{j}\epsilon_{isr}-\Lambda_{1}n_{s}\epsilon_{ijr}\right)\right]\\
		=&\Lambda_{1}\left(S\partial_{i}n_{j}+n_{s}\partial_{s}\left(\partial_{i}n_{j}\right)\right)+\Lambda_{1}\left(Sn_{i}B_{j}-B_{i}B_{j}+n_{i}n_{s}\partial_{s}B_{j}\right)\\
		-&\Lambda_{2}\left(n_{i}n_{j}\partial_{s}B_{s}+B_{s}\left(n_{i}\partial_{s}n_{j}+n_{j}\partial_{s}n_{i}\right)\right)\\
		+&\frac{1}{2}\Lambda_{2}\partial_{s}\left[Tn_{r}\left(n_{i}\epsilon_{jsr}+n_{j}\epsilon_{isr}\right)\right]-\frac{1}{2}\Lambda_{1}\partial_{s}\left(Tn_{r}n_{s}\epsilon_{ijr}\right)
	\end{split}
\end{equation*}
\begin{equation*}
	\begin{split}
		\partial_{k}\Gamma_{ij}^{k}=&\Lambda_{1}S\partial_{i}n_{j}+\Lambda_{1}\left[\cancel{B_{i}B_{j}}-\cancel{n_{i}\partial_{n}B_{j}}+\frac{1}{2}\partial_{n}S\left(\delta_{ij}-n_{i}n_{j}\right)+\frac{1}{2}S\left(n_{i}B_{j}+n_{j}B_{i}\right)+\frac{1}{2}\partial_{n}T\epsilon_{ijk}n_{k}-\frac{1}{2}T\epsilon_{ijk}B_{k}+\partial_{n}\Delta_{ij}\right]\\
		+&\Lambda_{1}Sn_{i}B_{j}-\cancel{\Lambda_{1}B_{i}B_{j}}+\cancel{\Lambda_{1}n_{i}\partial_{n}B_{j}}\\
		-&\Lambda_{2}n_{i}n_{j}\partial_{s}B_{s}-\Lambda_{2}\left[n_{i}\left(\frac{1}{2}SB_{j}-\frac{1}{2}TB_{s}\epsilon_{jsl}n_{l}+B_{s}\Delta_{js}\right)+n_{j}\left(\frac{1}{2}SB_{i}-\frac{1}{2}TB_{s}\epsilon_{isl}n_{l}+B_{s}\Delta_{is}\right)\right]\\
		+&\frac{1}{2}\Lambda_{2}\left[\partial_{s}Tn_{r}\left(n_{i}\epsilon_{jsr}+n_{j}\epsilon_{isr}\right)+T^{2}\left(3n_{i}n_{j}-\delta_{ij}\right)\right]\\
		+&\frac{1}{2}\Lambda_{2}\left[T\left(-n_{s}n_{r}\left(\epsilon_{jsr}B_{i}+\epsilon_{isr}B_{j}\right)+\epsilon_{jsr}n_{r}\Delta_{si}+\epsilon_{isr}n_{r}\Delta_{sj}\right)-Tn_{s}B_{r}\left(\epsilon_{jsr}n_{i}+\epsilon_{isr}n_{j}\right)\right]\\
		-&\frac{1}{2}\Lambda_{1}\left[n_{r}\epsilon_{ijr}\partial_{n}T+TS\epsilon_{ijr}n_{r}-TB_{r}\epsilon_{ijr}\right]\\
		=&\Lambda_{1}S\left[\cancel{-n_{i}B_{j}}+\frac{1}{2}S\left(\delta_{ij}-n_{i}n_{j}\right)+\cancel{\frac{1}{2}T\epsilon_{ijk}n_{k}}+\Delta_{ij}\right]\\
		+&\Lambda_{1}\left[\frac{1}{2}\partial_{n}S\left(\delta_{ij}-n_{i}n_{j}\right)+\frac{1}{2}S\left(n_{i}B_{j}+n_{j}B_{i}\right)+\cancel{\frac{1}{2}\partial_{n}T\epsilon_{ijk}n_{k}}-\cancel{\frac{1}{2}T\epsilon_{ijk}B_{k}}+\partial_{n}\Delta_{ij}+\cancel{Sn_{i}B_{j}}\right]\\
		-&\Lambda_{2}n_{i}n_{j}\partial_{s}B_{s}-\Lambda_{2}\left[n_{i}\left(\frac{1}{2}SB_{j}-\frac{1}{2}TB_{s}\epsilon_{jsl}n_{l}+B_{s}\Delta_{js}\right)+n_{j}\left(\frac{1}{2}SB_{i}-\frac{1}{2}TB_{s}\epsilon_{isl}n_{l}+B_{s}\Delta_{is}\right)\right]\\
		+&\frac{1}{2}\Lambda_{2}\left[\partial_{s}Tn_{r}\left(n_{i}\epsilon_{jsr}+n_{j}\epsilon_{isr}\right)+T^{2}\left(3n_{i}n_{j}-\delta_{ij}\right)\right]\\
		+&\frac{1}{2}\Lambda_{2}\left[T\left(\cancel{-n_{s}n_{r}\left(\epsilon_{jsr}B_{i}+\epsilon_{isr}B_{j}\right)}+\epsilon_{jsr}n_{r}\Delta_{si}+\epsilon_{isr}n_{r}\Delta_{sj}\right)-Tn_{s}B_{r}\left(\epsilon_{jsr}n_{i}+\epsilon_{isr}n_{j}\right)\right]\\
		-&\frac{1}{2}\Lambda_{1}\left[\cancel{n_{r}\epsilon_{ijr}\partial_{n}T}+\cancel{TS\epsilon_{ijr}n_{r}}-\cancel{T\epsilon_{ijr}B_{r}}\right]
	\end{split}
\end{equation*}
\begin{equation*}
	\begin{split}
		\Gamma_{ik}^{r}\Gamma_{rj}^{k}=&-\left(\Lambda_{1}-\Lambda_{2}\right)\left[\frac{1}{2}Tn_{l}\left(\epsilon_{jkl}\Delta_{ik}+\epsilon_{ikl}\Delta_{jk}\right)-\frac{1}{2}S\left(n_{j}B_{i}+n_{i}B_{j}\right)-\left(n_{j}\Delta_{ik}B_{k}+n_{i}\Delta_{kj}B_{k}\right)\right]\\
		-&\Lambda_{2}^{2}\frac{1}{2}T^{2}n_{i}n_{j}
	\end{split}
\end{equation*}
Summing it all up gives us
\begin{equation*}
	\begin{split}
		R_{ij}=&\partial_{k}\Gamma_{ij}^{k}-\Gamma_{ik}^{r}\Gamma_{rj}^{k}\\
		=&\Lambda_{1}\left[\frac{1}{2}\left(S^{2}+\partial_{n}S\right)\left(\delta_{ij}-n_{i}n_{j}\right)+S\Delta_{ij}+\partial_{n}\Delta_{ij}\right]\\
		+&\frac{1}{2}\left(\Lambda_{1}-\Lambda_{2}\right)\cancel{S\left(n_{i}B_{j}+n_{j}B_{i}\right)}\\
		-&\Lambda_{2}n_{i}n_{j}\partial_{s}B_{s}-\Lambda_{2}n_{i}B_{s}\Delta_{js}-\Lambda_{2}n_{j}B_{s}\Delta_{is}\\
		+&\frac{1}{2}\Lambda_{2}\left[\partial_{s}Tn_{r}\left(n_{i}\epsilon_{jsr}+n_{j}\epsilon_{isr}\right)+T^{2}\left(3n_{i}n_{j}-\delta_{ij}\right)\right]\\
		+&\frac{1}{2}\Lambda_{2}T\left(\epsilon_{jsr}n_{r}\Delta_{si}+\epsilon_{isr}n_{r}\Delta_{sj}\right)+\Lambda_{2}T\left(n_{i}B_{s}n_{l}\epsilon_{jsl}+n_{j}n_{l}B_{s}\epsilon_{isl}\right)\\
		+&\left(\Lambda_{1}-\Lambda_{2}\right)\left[\frac{1}{2}Tn_{l}\left(\epsilon_{jkl}\Delta_{ik}+\epsilon_{ikl}\Delta_{jk}\right)-\cancel{\frac{1}{2}S\left(n_{j}B_{i}+n_{i}B_{j}\right)}-\left(n_{j}\Delta_{ik}B_{k}+n_{i}\Delta_{kj}B_{k}\right)\right]\\
		+&\Lambda_{2}^{2}\frac{1}{2}T^{2}n_{i}n_{j}\\
		=&\Lambda_{1}\left[\frac{1}{2}\left(S^{2}+\partial_{n}S\right)\left(\delta_{ij}-n_{i}n_{j}\right)+S\Delta_{ij}+\partial_{n}\Delta_{ij}\right]\\
		+&\underset{-\Lambda_{1}\Lambda_{2}}{\underbrace{\left(\Lambda_{1}-\Lambda_{2}\right)}}\left[\frac{1}{2}Tn_{l}\left(\epsilon_{jkl}\Delta_{ik}+\epsilon_{ikl}\Delta_{jk}\right)-\left(n_{j}\Delta_{ik}B_{k}+n_{i}\Delta_{kj}B_{k}\right)\right]\\
		-&\Lambda_{2}n_{i}B_{s}\Delta_{js}-\Lambda_{2}n_{j}B_{s}\Delta_{is}\\
		+&\Lambda_{2}\partial_{s}Tn_{r}\frac{n_{i}\epsilon_{jsr}+n_{j}\epsilon_{isr}}{2}+\frac{1}{2}\Lambda_{2}T^{2}\left(3n_{i}n_{j}-\delta_{ij}\right)\\
		+&\Lambda_{2}Tn_{r}\frac{\epsilon_{jsr}\Delta_{si}+\epsilon_{isr}\Delta_{sj}}{2}+\Lambda_{2}T\left(n_{i}B_{s}n_{l}\epsilon_{jsl}+n_{j}n_{l}B_{s}\epsilon_{isl}\right)\\
		+&\Lambda_{2}^{2}\frac{1}{2}T^{2}n_{i}n_{j}-\Lambda_{2}n_{i}n_{j}\partial_{s}B_{s}
	\end{split}
\end{equation*}
Overall we get
\begin{equation*} \label{eqn: Ricci tensor result}
	\begin{split}
		R_{ij}=&\Lambda_{1}\left[\frac{1}{2}\left(S^{2}+\partial_{n}S\right)\left(\delta_{ij}-n_{i}n_{j}\right)+S\Delta_{ij}+\partial_{n}\Delta_{ij}\right]\\
		+&\left(\Lambda_{1}-\Lambda_{2}\right)\left[\frac{1}{2}Tn_{l}\left(\epsilon_{jkl}\Delta_{ik}+\epsilon_{ikl}\Delta_{jk}\right)-\left(n_{j}\Delta_{ik}B_{k}+n_{i}\Delta_{kj}B_{k}\right)\right]\\
		-&\Lambda_{2}\left[n_{i}n_{j}\partial_{s}B_{s}+B_{s}\left(n_{i}\Delta_{js}+n_{j}\Delta_{is}\right)\right]\\
		+&\Lambda_{2}\left[\partial_{s}Tn_{r}\frac{n_{i}\epsilon_{jsr}+n_{j}\epsilon_{isr}}{2}+\frac{1}{2}T^{2}\left(3n_{i}n_{j}-\delta_{ij}\right)\right]\\
		+&\Lambda_{2}\left[Tn_{r}\frac{\epsilon_{jsr}\Delta_{si}+\epsilon_{isr}\Delta_{sj}}{2}+TB_{s}\left(n_{i}n_{l}\epsilon_{jsl}+n_{j}n_{l}\epsilon_{isl}\right)\right]\\
		+&\frac{1}{2}\Lambda_{2}^{2}T^{2}n_{i}n_{j}
	\end{split}
\end{equation*}

Further simplifications yields the result:
\begin{equation*}
	\begin{split} \label{eqn: ricci tensor, biaxial w/o projection}
		R_{ij}=&-\Lambda_{2}n_{i}n_{j}\left[\nabla\cdot\boldsymbol{B}-\left(1+\frac{1}{2}\Lambda_{2}\right)T^{2}\right]\\+&\frac{\Lambda_{1}}{2}\left(\delta_{ij}-n_{i}n_{j}\right)\left[\nabla\cdot\left(S{\bf n}\right)-\frac{\Lambda_{2}}{\Lambda_{1}}T^{2}\right]\\+&\Lambda_{1}\left[\partial_{n}\Delta_{ij}+S\Delta_{ij}+Tn_{l}\epsilon_{jsl}\Delta_{is}-B_{k}\left(n_{j}\Delta_{ik}+n_{i}\Delta_{jk}\right)\right]\\+&\Lambda_{2}\left(n_{i}\epsilon_{jsl}n_{l}+n_{j}\epsilon_{isl}n_{l}\right)\left[\frac{1}{2}\partial_{s}T+TB_{s}\right]
	\end{split}
\end{equation*}
where we denoted $\Lambda_1 =\frac{\lambda_1^2-\lambda_2^2}{\lambda_1^2}$ and $\Lambda_2 =\frac{\lambda_1^2-\lambda_2^2}{\lambda_2^2}$.
Apparently, the Ricci tensor has a composite structure of two scalars, a vector and a tensor. In order to proceed we need to understand the behavior of the tensor $\Delta_{ij}$. 
Any variation in this tensor can be understood as either one of 3 categories: changes in the magnitude of the tensor ($\Delta=\sqrt{\Delta_1^2+\Delta_2^2}$), rotation around the ${\bf n}$ axis, and orientation change due to bending of the director. 

In order to analyze these modes of variations we first need to establish some properties of $\Delta_{ij}$. First we study its variation along $\partial_n$ using its orthogonality to the director.
We use the identity from eq.~\eqref{eqn: Delta conservation} that $n_{i}\partial_{n}\Delta_{ij}=B_{i}\Delta_{ij}$. From this we can infer that
\begin{equation*}
	n_{i}n_{j}\partial_{n}\Delta_{ij}=n_{j}B_{i}\Delta_{ij}=0
\end{equation*}
This allows us to explicitly write the projection $\mathcal{P}_{\perp}\left[\partial_{n}\Delta_{ij}\right]$ as
\begin{equation*}
	\begin{split}
		\mathcal{P}_{\perp}\left[\partial_{n}\Delta_{ij}\right]\equiv&\left(\delta_{is}-n_{i}n_{s}\right)\left(\delta_{jr}-n_{j}n_{r}\right)\partial_{n}\Delta_{rs}\\=&\partial_{n}\Delta_{ij}-n_{j}n_{r}\partial_{n}\Delta_{ir}-n_{i}n_{s}\partial_{n}\Delta_{js}+n_{i}n_{j}\cancel{n_{r}n_{s}\partial_{n}\Delta_{rs}}\\=&\partial_{n}\Delta_{ij}-n_{j}B_{k}\Delta_{ik}-n_{i}B_{k}\Delta_{jk}
	\end{split}
\end{equation*}
therefore we can write the third part of the Ricci tensor as a projection onto the subspace perpendicular to ${\bf n}$:
\begin{equation*}
	\begin{split}
		R_{ij}=&-\Lambda_{2}n_{i}n_{j}\left[\nabla\cdot\boldsymbol{B}-\left(1+\frac{1}{2}\Lambda_{2}\right)T^{2}\right]\\+&\frac{\Lambda_{1}}{2}\left(\delta_{ij}-n_{i}n_{j}\right)\left[\nabla\cdot\left(S{\bf n}\right)-\frac{\Lambda_{2}}{\Lambda_{1}}T^{2}\right]\\+&\Lambda_{1}\left(\delta_{jr}-n_{j}n_{r}\right)\left(\delta_{is}-n_{i}n_{s}\right)\left[\partial_{n}\Delta_{sr}+S\Delta_{sr}+Tn_{l}\epsilon_{rkl}\Delta_{sk}\right]\\+&\Lambda_{2}\left(n_{i}\epsilon_{jsl}n_{l}+n_{j}\epsilon_{isl}n_{l}\right)\left[\frac{1}{2}\partial_{s}T+TB_{s}\right]
	\end{split}
\end{equation*}

\newpage
\section{Refining compatibility conditions} \label{sec: refining compatibility conditions}

The resulting eqs.~\eqref{eqn: Ricci Components} can be simplified into the set of equations \eqref{eqn: Simplified Ricci} through the following: 

First we derive eq.~\eqref{eqn: Simplified Ricci 1}: 
The equation $\Lambda^2 T^2=\nabla \cdot (S{\bf n})$ is equivalent to eq.~\eqref{eqn: Ricci Components 2}. Then we subtract eq.~\eqref{eqn: Ricci Components 1} from eq.~\eqref{eqn: Ricci Components 2} to obtain
\begin{equation*}
	\nabla \cdot (S {\bf n} + \boldsymbol{B})-T^2 \left( \Lambda^2+\frac{\Lambda^2+1}{2} \right)=0
\end{equation*}
Identifying $\sigma =\nabla \cdot (S {\bf n} + \boldsymbol{B})$ finishes this derivation.

To derive eqs.~\eqref{eqn: Simplified Ricci 2}, \eqref{eqn: Simplified Ricci 3} we first make a few observations with regard to the $\Delta$-mode which is symmetric, traceless, and perpendicular to ${\bf n}$, namely
\begin{equation*}
	\Delta_{ij}=\Delta_{ji},\quad \Delta_{ii}=0,\quad \textrm{and} \quad n_i\Delta_{ij}=0.
\end{equation*}
These conditions hold at every point, thus their derivative along ${\bf n}$ vanishes:
\begin{equation}\label{eqn: Delta conservation}
	\partial_n\Delta_{ij}=\partial_n\Delta_{ji},\quad \partial_n\Delta_{ii}=0,\quad \textrm{and} \quad \partial_n(n_i\Delta_{ij})=-B_i\Delta_{ij}+n_i\partial_n\Delta_{ij}=0.
\end{equation}

We now turn to write $\partial_n\Delta_{ij}$ as a sum of its perpendicular-to-${\bf n}$ and parallel-to-${\bf n}$ components:
\begin{equation*}
	\partial_n\Delta_{ij}=\mathcal{P}_\perp(\partial_n\Delta_{ij})+n_i \partial_n \Delta_{kj}n_k+\partial_n\Delta_{ik}n_k n_j.
\end{equation*}
The parallel terms originate from the bend (which changes the plane in which the $\Delta$-mode lives and are given explicitly by the last equation in \eqref{eqn: Delta conservation}. We are nonetheless interested in the perpendicular component $\mathcal{P}_\perp(\partial_n\Delta_{ij})$, which, by the first two equations in \eqref{eqn: Delta conservation}, is also symmetric and traceless. It is therefore spanned by $\Delta_{ij}$ and its (symmetric traceless) $90^\circ$ rotation around ${\bf n}$, $\epsilon_{ikl}n_k\Delta_{lj}$. Thus:
\begin{equation*}
	\begin{split}
		\mathcal{P}_\perp(\partial_n\Delta_{ij})&=\left(\frac{1}{2\Delta^2}\Delta_{kl}\partial_n\Delta_{kl}\right)\Delta_{ij}+\left(\frac{1}{2\Delta^2}\epsilon_{mkl}n_k\Delta_{lp}\partial_n\Delta_{mp}\right)\epsilon_{ikl}n_k\Delta_{lj}\\
		&=\frac{\partial_n\Delta}{\Delta}\Delta_{ij}+T_{\Delta}\epsilon_{ikl}n_k\Delta_{lj}
	\end{split}
\end{equation*}
where $T_{\Delta}$ is the $\Delta$-twist, the rate by which $\Delta_{ij}$ rotates within the plane perpendicular to the director when moving along the director. It takes the explicit form
\begin{equation*}
	T_{\Delta}=\frac{1}{2\Delta^2}\epsilon_{mkl}n_k\Delta_{lp}\partial_n\Delta_{mp}=2\epsilon_{mkl}n_ke_{l}\partial_ne_{m},
\end{equation*}
where $\hat{e}$ is one of the unit eigenvectors of $\boldsymbol{\Delta}$ perpendicular to ${\bf n}$.

\add{
\newpage
\section{Twist must vanish in a compatible bulk} \label{sec: appendix twist}

Here we prove that if the compatibility conditions \eqref{eqn: Ricci Components} hold in an open 3D region and $\Lambda \neq 1$ (with $\Lambda \neq 0$), then $T \equiv 0$ throughout that region. Consequently, Eq.~\eqref{eqn: Simplified Ricci 1} reduces to $\sigma = 0$, and temperature-selective compatibility cannot be realized in a 3D volume. To help the reader, this appendix is accompanied by a Wolfram Mathematica notebook which provides a direct symbolic verification of all the equations herein. 

We denote the left-hand sides of Eqs.~\eqref{eqn: Ricci Components} by
\begin{equation}
	\begin{split}
	    E^b &= \partial_k(n_k S) - \Lambda^2 T^2, \\
		E^a &= -\partial_k B_k + \frac{\Lambda^2+1}{2} T^2,\\
		E_{ij}^c &= \mathcal{P}_{ir}\mathcal{P}_{js}(\partial_n \Delta_{rs} + S\Delta_{rs} - T\tilde{\Delta}_{rs}), \\
		E_i^d &= \mathcal{P}_{ik}\partial_k T + 2T B_i,
	\end{split}\label{eqn: E def}
\end{equation}
where $\mathcal{P}_{ij} \equiv \delta_{ij} - n_i n_j$ is the projection operator onto the plane perpendicular to ${\bf n}$, and a tilde denotes a $90^\circ$ rotation in this plane around ${\bf n}$: $\tilde{v}_i \equiv \epsilon_{ikl}n_k v_l$ for a vector and $\tilde{\Delta}_{ij} \equiv \epsilon_{ikl}n_k \Delta_{lj}$ for $\boldsymbol{\Delta}$. We call any linear combination of $E^a, E^b, E_{ij}^c, E_i^d$ and their spatial derivatives, with coefficients constructed from ${\bf n}$ and its derivatives, a \emph{residue}. By definition, any residue vanishes identically in any open region where Eqs.~\eqref{eqn: Ricci Components} hold. For example, subtracting $E^a$ from $E^b$ and using the saddle-splay identity \eqref{def: saddle-splay}, $\partial_k(n_k S + B_k) = \frac{1}{2}S^2 + \frac{1}{2}T^2 - 2\Delta^2$, yields the residue
\begin{equation}
    F \equiv E^b - E^a = \frac{1}{2}S^2 - 2\Delta^2 - \frac{3}{2}\Lambda^2 T^2. \label{eqn: residue F}
\end{equation}
Wherever Eqs.~\eqref{eqn: Ricci Components} hold, $F = 0$ and thus $S^2 = 4\Delta^2 + 3\Lambda^2 T^2$.

Because $S, \boldsymbol{B}, T,$ and $\boldsymbol{\Delta}$ are all defined in terms of $\partial_i n_j$ via Eq.~\eqref{eqn: definition gradient n}, the commutation of partial derivatives ($\partial_i\partial_j n_k = \partial_j\partial_i n_k$) imposes differential identities among them. Using Eq.~\eqref{eqn: definition gradient n} to evaluate $\partial_k \Delta_{kj}$ and $\epsilon_{ijk}n_i \partial_j E_k^d$, a direct calculation gives the two first-order identities
\begin{align}
    2\mathcal{P}_{ij}\partial_k \Delta_{kj} &= \mathcal{P}_{ij}\partial_j S + 2\Delta_{ij}B_j + \tilde{E}_i^d, \label{eqn: diff id 1} \\
    -3T\partial_n T - 2ST^2 &= \epsilon_{ijk}n_i \partial_j E_k^d + 2\tilde{B}_i E_i^d \equiv \mathcal{R}_T, \label{eqn: diff id 2}
\end{align}
where $\tilde{E}_i^d$ and $\mathcal{R}_T$ are residues.

At higher order, one can form a linear combination of the first and second derivatives of $E^a, E^b, E_{ij}^c, E_i^d$ in which all third derivatives of ${\bf n}$ cancel, and use the integrability of $\partial_i(T^2) = -4T^2(\frac{1}{3}Sn_i + B_i) + 2T E_i^d - \frac{2}{3}\mathcal{R}_T n_i$ (obtained by combining $E_i^d$ in Eq.~\eqref{eqn: E def} with $\mathcal{R}_T$ in Eq.~\eqref{eqn: diff id 2}) to eliminate all remaining second derivatives of ${\bf n}$. Carrying out this elimination without division yields the exact identity
\begin{equation}
    T^4 \mathcal{S} = \mathcal{R}, \label{eqn: T4 S eq R}
\end{equation}
where $\mathcal{R}$ is a residue (listed explicitly below) and $\mathcal{S}$ is a sum of squares:
\begin{equation}
    \mathcal{S} = \frac{1}{2}|(S\mathcal{P}_{ij} - 2\Delta_{ij})K_j|^2 + T^2\left[\frac{3}{2}\Lambda^2 K_i K_i + 48\Lambda^2 \Delta^2 B_i B_i + 32(\Lambda^2+1)\Delta^4\right], \label{eqn: S sum of squares}
\end{equation}
with $K_i \equiv \mathcal{P}_{ij}\partial_j S + 2S B_i + 2\Delta_{ij}B_j$ and $|v_i|^2 \equiv v_i v_i$. Equation~\eqref{eqn: T4 S eq R} holds for any smooth unit director field ${\bf n}$ and can be verified by direct expansion using Eq.~\eqref{eqn: definition gradient n}.

\subsection*{Proof that $T \equiv 0$}

Suppose Eqs.~\eqref{eqn: Ricci Components} hold in an open region, and let $U$ be the open subset of that region where $T \neq 0$. All residues vanish identically in $U$.
\begin{enumerate}
    \item In $U$, $\mathcal{R} = 0$ and $T \neq 0$, so Eq.~\eqref{eqn: T4 S eq R} gives $\mathcal{S} = 0$. Since every term in Eq.~\eqref{eqn: S sum of squares} is non-negative, each must vanish separately. Because $T \neq 0$ in $U$ and $\Lambda \neq 0$, the terms in the brackets of Eq.~\eqref{eqn: S sum of squares} give $\Delta \equiv 0$ (meaning $\Delta_{ij} \equiv 0$, as $\Delta_{ij}\Delta_{ij} = 2\Delta^2$) and $K_i \equiv 0$ in $U$.
    \item With $\Delta_{ij} \equiv 0$ and $\tilde{E}_i^d = 0$, the first identity in Eq.~\eqref{eqn: diff id 1} reduces to $\mathcal{P}_{ij}\partial_j S = 0$. Substituting $\Delta_{ij} = 0$ and $\mathcal{P}_{ij}\partial_j S = 0$ into the definition of $K_i$ gives $0 = K_i = 2SB_i$. Meanwhile, Eq.~\eqref{eqn: residue F} with $F = 0$ and $\Delta = 0$ gives $S^2 = 3\Lambda^2 T^2 \neq 0$, so $S \neq 0$ and therefore $B_i \equiv 0$ in $U$.
    \item With $E_i^d = 0$ and $B_i = 0$, Eq.~\eqref{eqn: E def} gives $\mathcal{P}_{ik}\partial_k T = 0$, while the second identity in Eq.~\eqref{eqn: diff id 2} gives $\partial_n T = -\frac{2}{3}ST$. Thus the full gradient of $T$ is $\partial_i T = (\mathcal{P}_{ik}\partial_k T) + n_i \partial_n T = -\frac{2}{3}ST n_i$. Since a gradient is curl-free, $0 = \epsilon_{ijk}\partial_j(\partial_k T) = -\frac{2}{3}\epsilon_{ijk}\partial_j(ST n_k)$. Contracting with $n_i$ gives
    \begin{equation*}
        0 = n_i \epsilon_{ijk}\partial_j(ST n_k) = ST(n_i \epsilon_{ijk}\partial_j n_k) = ST^2,
    \end{equation*}
    which contradicts $S \neq 0$ and $T \neq 0$ in $U$.
\end{enumerate}
Hence the open set $U$ must be empty, meaning $T \equiv 0$ throughout the region.

\subsection*{Explicit residue}

To write the residue $\mathcal R$ in eq.~\eqref{eqn: T4 S eq R} compactly, we first define the two vector residues
\begin{equation*}
	\mathcal R_i\equiv P_{ij}\epsilon_{jkl}\Big(\frac12\partial_k(TE^d_l)-\frac16n_l\partial_k\mathcal R_T-2TE^d_kB_l-\frac23STE^d_kn_l+\frac56\mathcal R_Tn_kB_l\Big),\qquad
	G_i\equiv P_{ij}\partial_jF-\tilde\Delta_{ij}E^d_j+3\Lambda^2TE^d_i,
	\label{eq:RiGi}
\end{equation*}
where $\mathcal R_i$ satisfies $T^2P_{ij}\epsilon_{jkl}\partial_k\big(\frac13Sn_l+B_l\big)=\mathcal R_i$. In terms of $F$, $\mathcal R_T$, $\mathcal R_i$, and $G_i$, the residue $\mathcal R$ in eq.~\eqref{eqn: T4 S eq R} is
\begin{equation*}
	\begin{aligned}
		\mathcal R={}&6\Delta^2\Big(2\tilde\Delta_{ij}T^2\partial_j\mathcal R_i-S\,T^2\epsilon_{ijk}n_i\partial_j\mathcal R_k+\big[3T^2(2\tilde\Delta_{ij}B_j-S\tilde B_i)-2\tilde\Delta_{ij}\partial_j(T^2)+S\,\epsilon_{ijk}n_j\partial_k(T^2)\big]\mathcal R_i\Big)\\
		&-2T^4\Delta^2\Big(P_{ij}\partial_i\partial_jF+3S\,\partial_nE^a-6\Delta_{ij}\partial_nE^c_{ij}+3\Lambda^2T\,\partial_iE^d_i-2\tilde\Delta_{ij}\partial_jE^d_i+S\mathcal R_T\Big)\\
		&+T^4\Big[8B_i(SP_{ij}-\Delta_{ij})K_j-K_iK_i-16\Delta^2B_iB_i-16\Delta^4-2(2+5\Lambda^2)T^2\Delta^2-(16B_iB_i+4\Delta^2)F-8B_iG_i\Big]F\\
		&+2T^4\Delta^2\big(8\Delta^2-9\Lambda^2T^2-2F\big)E^a-4T^4\Delta^2\big(2S\Delta_{ij}+T\tilde\Delta_{ij}\big)E^c_{ij}\\
		&+2T^4\Delta^2\big(21\Lambda^2TB_i+\epsilon_{ikl}n_k\partial_lS-3S\tilde B_i-3\tilde\Delta_{ij}B_j\big)E^d_i\\
		&-(1+6\Lambda^2)T^4\Delta^2E^d_iE^d_i+T^4\big[2(SP_{ij}-\Delta_{ij})K_j-6\Delta^2B_i-G_i\big]G_i.
	\end{aligned}
	\label{eq:Rtotal}
\end{equation*}
}

\end{document}